# CTA telescopes as deep-space lasercom ground receivers

Alberto Carrasco-Casado
Space Communication Systems Laboratory
National Institute of Information and Communications Technology
4-2-1 Nukui-Kitamachi, Koganei, Tokyo 184-8795, Japan
alberto@nict.go.jp

José Manuel Sánchez-Pena, Ricardo Vergaz
Electronic Technology Department, Carlos III University of Madrid
Av. de la Universidad 30, Madrid 28911, Spain

*Abstract*—The amount of scientific data to be transmitted from deep-space probes is currently very limited due to RF-communications constraints. Free-space optical communication promises to alleviate this bottleneck as this technology makes it possible to increase the data rate while reducing the weight, mass and power of communication onboard equipment. Nevertheless, further improvements are needed to optimize the power delivery from the spacecraft to the Earth. This has been also a major issue in RF communications, where the main strategy has been to increase the aperture of ground terminals. Free-space optical communications can also be benefited from this strategy, as it shares the same limitation with RF, i.e. the low power received on the Earth. However, the cost of big telescopes increases exponentially with their aperture, being much bigger than the cost of big antennas. Therefore, new ideas are required to maximize the aperture-to-cost ratio. This work explores the feasibility of using telescopes of the future Cherenkov Telescope Array as optical-communication ground stations. Cherenkov telescopes are used for gamma-ray astronomy, yet they are optical telescopes with some special characteristics. Ground-based gamma-ray astronomy has the same received-power limitation as deep-space lasercom, hence Cherenkov telescopes are designed to maximize the receiver's aperture, reaching up to 30-m diameters, at a minimum cost with some relaxed requirements. Discussions on the critical issues of the reutilization, as well as possible adaptations of the telescopes to optimize them for communications, are presented. Telescopes simulations and numerical computations of several link budgets applied to different worst-case scenarios are discussed, concluding that the proposal is technically feasible and would bring important cost reductions as well as performance improvements compared to current designs for deep-space optical ground stations.

## I. INTRODUCTION

Payloads in deep-space missions are continually increasing the amount of information collected during their lifetimes. The result is that bigger and bigger amounts of data need to be transmitted back to Earth, and current radiofrequency (RF) technology has reached a bottleneck, limiting the scientific outcome of current missions and threatening the development of future manned missions. To solve this, optical frequencies have been studied and applied [1] to increase the bandwidth and to reduce the volume, mass and power needs at the same time. One of the main advantages of shorter wavelengths is the narrow divergence of the laser beams. For example, a communication link from Mars would allow a reduction of the footprint reaching the Earth from ~1000 in RF to ~0.1 of the Earth diameter using optical wavelengths.

A natural need of a Free-Space Optical Communication (FSOC) link, especially from deep-space, is the use of a ground terminal, i.e., a telescope, with a big aperture to overcome the very low power received. Nevertheless, unlike conventional astronomical telescopes, in FSOC there is no need to obtain images, thus the requirements are not so strong in terms of optical quality. This is one of the reasons to study the use of other kind of telescopes. This paper completes an initial proposal from the authors about the use of IACTs (Imaging Atmospheric Cherenkov Telescopes) as optical ground terminals in deep-space FSOC [2].

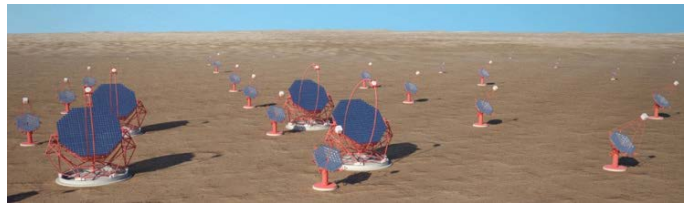

Fig. 1. Artistic illustration of the Cherenkov Telescope Array [3].

## II. PROBLEM AND SOLUTION PROCEDURE

Cherenkov radiation is produced in the atmosphere when a cosmic ray or a gamma ray interacts with the molecules of the atmospheric upper layers. As a result, visible photons are produced, and IACTs are designed to detect them in the ground using a big reflective surface. These telescopes observe the atmosphere at a height of ~10 km, where on average the Cherenkov radiation is produced. The Cherenkov Telescope Array (CTA) is a 200 M€ international collaboration involving 31 countries to build over a hundred of big segmented telescopes (fig. 1) in the next years, deployed in two observatories, one in each Earth's hemisphere to allow full coverage of the sky [3]. Recently, the CTA Resource Board decided to host CTA-South in the ESO Paranal grounds (Chile), and CTA-North in Roque de los Muchachos Observatory (La Palma, Spain) [4]. Several telescope prototypes have already been tested and the final telescopes will be built in the next few years with the goal of being fully operational in 2020.

The CTA observatory will consist of about 100 telescopes on the southern site and about 20 telescopes on the northern site. In each observatory, there will be three types of different telescopes (fig. 2):

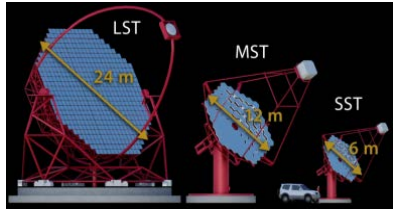

Fig. 2. Diagram of the three types of different CTA telescopes [3].

- SST (Small Size Telescopes): with 6-m diameter and 10º field of view (FoV). A pool of dozens of them, separated by at least 70 m, is projected, thus a massive production will be carried out.

- MST (Medium Size Telescopes): with 12-m diameter and 6º to 8º FoV. Tens of them with 100-m spacing are foreseen, being the most similar to current IACTs.

- LST (Large Size Telescopes): with 24-m diameter and 4º to 5º FoV, 100-m spacing and around 4 units in each observatory. The first prototype of this kind is currently under construction in La Palma, Spain.

The authors of this paper presented the original idea of using IACT telescopes for deep-space FSOC in a previous paper [2]. The motivations were explored back then: the big apertures of the telescopes, their native operation as an array, their low costs due to their relaxed requirements, the ideal sky-related conditions, the fast tracking of the telescopes, a suitable communications network infrastructure, etc. Now, the proposal is studied in more depth and extended to CTA telescopes. Data on the telescope designs has been taken from those planned under current CTA consideration.

In the next sections, a deep insight into the main differences of gamma-ray and communication telescopes will be carried out, focusing on the telescopes from the CTA project. Originally, this proposal is aimed to the direct reutilization of one or more CTA telescopes for their exclusive use as communication receivers. However, a different approach could also be taken into consideration, i.e. the shared operation for astronomy and communications. The direct reutilization of CTA telescopes, with minor modifications, is proposed, although a brief discussion on possible improvements to optimize their performance for FSOC will be made.

In this study, reflectance/transmittance measurements were performed using a Lambda 900 spectrometer by Perkin Elmer. OSLO (Optics Software for Layout and Optimization) software from Lambda Research was used for the optical models and simulations, allowing the computation of optical performance of the telescopes. MODTRAN (MODerate resolution atmospheric TRANsmission) software from Spectral Sciences was used as the standard tool to retrieve atmospheric transmission and sky radiance, in order to compute received noise in simulations at different scenarios. Matlab from MathWorks was used for the rest of computations. Lastly, the recommendations from Optical Link Study Group (OLSG), a subcommittee of the Interagency Operations Advisory Group (IOAG), co-chaired by ESA and NASA and in charge of the international standardization of FSOC, were followed in the simulations.

III. RESULTS AND DISCUSSION

In order to fulfill the requirements of FSOC, some IACTs items must be carefully studied. The reflectivity of the mirrors at the desired wavelength, the different position of the image plane and the focal plane, the pointing and tracking of the telescope gimbal, and a brief discussion of the optimization of the detector size were already made in [2]. Hereafter, updates on the reflectivity and focusing with new CTA solutions as well as FoV limitations for the SNR (Signal to Noise Ratio) in this kind of telescopes will be explored. After that, a number of telescope simulations will be performed to validate their proper operation for FSOC. With these results, a summary of the link budgets that can be achieved for different scenarios will be made, concluding with a brief analysis of the costs.

*A. Mirrors of CTA telescopes*

The first item to be explored is the performance of the mirrors at FSOC wavelength. Several techniques are being proposed for manufacturing CTA segments of primary mirrors. Due to the foreseen massive production, the spherical mirror profiles and the moderate optical quality required, most techniques are based on replica to minimize costs (~2000 €/m$^2$), where glass is conformed using a molding based on a honeycomb structure [5], but the curvature radius that this technique can achieve is too low for most SST and probably so for MST as well. Thus, more traditional polishing techniques with diamond in aluminum (~2500 €/m$^2$) will be also examined. Finally, to avoid degradation in aluminum layers and increase reflectivity, interference dielectric multilayer designs are also under study [6]. In these mirrors, the reflectivity can be adjusted in narrow spectral bands. The main question to be solved is whether the mirrors can be directly used in FSOC wavelength (1550 nm), provided that they are designed for Cherenkov radiation band (visible light).

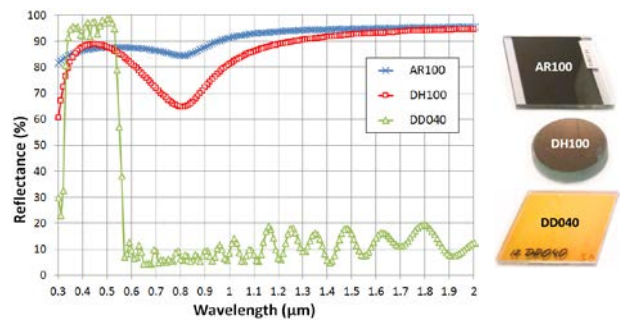

Fig. 3. Measured reflectance of the mirrors types considered for CTA.

Reflectance measurements have been carried out over different mirrors currently intended to be used in CTA [2]. These measurements showed that the reflectance at FSOC wavelengths is even higher than in the Cherenkov region for current IACT mirrors, reaching over 90% at 1550 nm. Fig. 3 shows the measured reflectance of several replica-based samples supplied by the CTA consortium: AR100 is made up of an aluminum layer and a quartz coating (Al+SiO2), DH100 is based on an aluminum layer with dielectric multilayer coating (SiO2+HfO2+SiO2), and DD040 is a pure dielectric multilayer mirror. AR100 reaches 94.9% at 1550 nm (around 85% for Cherenkov range), and DH100 reaches 93.0% (88.8% at Cherenkov range). The remarkable interference effect of DD040 reaches a 98.8% at Cherenkov wavelengths, although a poor 7.3% at 1550 nm. In principle, this could be a problem, since this technique is only starting to be spread. However, pure dielectric mirrors can be tuned to maximize their reflectance in a different wavelength during the manufacture process with no additional cost [7]. Furthermore, mirrors of this kind were installed in Hess telescope (28-m IACT located in Namibia) in 2012 showing a nearly 100% reflectance at 1550 nm, although for reasons not related to communications.

### B. Focusing an IACT

In [2], the required detector position displacement $\varepsilon$ in order to focus at infinity (as a FSOC receiver) instead of at a point 10 km high in the atmosphere (as IACTs are designed to operate, in order to detect the Cherenkov radiation), was studied for the MAGIC (Major Atmospheric Gamma Imaging Cherenkov Telescope) telescope. The same equations of [2], applied to CTA, give $\varepsilon$ values ranging from 1.4 cm (SST) to 9 cm (LST) (see table I). In the case of MAGIC II, the camera can be shifted as far as 30 cm in real operations for focusing and maintenance [8], so this is a normal feature in IACTs. Furthermore, when adapting an IACT as a FSOC receiver, the big and heavy cameras should be replaced by a simpler system based on a single photodetector, leaving enough space to allocate any optical setup with no significant constraints.

TABLE I. DETECTOR DISPLACEMENT $\varepsilon$ IN IACTS FOR FSOC OPERATION.

| Telescope | Diameter D (m) | f/D | Focal length f (m) | ε (cm) |
|---|---|---|---|---|
| MAGIC | 17 | 1 | 17 | 2.9 |
| CTA-SST | 6 | 0.5 | 12 | 1.4 |
| CTA-MST | 12 | 1 / 0.75 | 16 | 2.5 |
| CTA-LST | 24 | 1.25 | 30 | 9 |

### C. Field of View and Background Noise

In [2], a discussion about the relation between the FoV and the SNR was made for MAGIC-II telescope. A deeper discussion needs to be made regarding the concepts of SNR, FoV and their relation with the optical resolution of the CTA telescopes. The optical resolution is characterized by the Point Spread Function (PSF), which determines the spatial distribution of the radiation received from a point source at infinity, i.e., the radiation coming from a space probe received at the focal plane of the telescope. The FoV $\theta_{FOV}$ is the angular spread in the object plane projected on the image plane of the telescope. It is also a function of the detector size (in this work, a circular shape is assumed), given by eq. (1).

$$\theta_{FOV} = 2\arctan(d/2f) \quad (1)$$

$d$ being the diameter of the detector, and $f$ the focal length of the telescope. The multi-pixel camera (made up by up to 2500 photomultiplier tubes in the case of LST) designed for CTA telescopes should be replaced by a simpler system based on a single photodetector, with an important reduction in the detector size, and therefore in the FOV.

Cherenkov telescopes operate during dark nights only. However, FSOC telescopes must support also daylight operation. In this case, the sunlight, which is the main source of background radiation due to the scattering caused by atmospheric gases (Rayleigh) and particles (Mie), reaches the detector as a noisy signal, hence with a decrease in the SNR. In this work, a worst case for each link scenario will be studied, and this spurious light will lead to a detected noise power $N_S$. According to eq. (2), this power depends on the sky radiance $L(\lambda,\theta,\varphi)$ (which in turn depends on wavelength $\lambda$, zenith angle $\theta$, and the angle $\varphi$ between the telescope, the target and the Sun), the receiver's aperture area $A_r$, the two-dimensional FoV angle $\Omega_{FOV}$, and the spectral bandwidth of the signal $\Delta\lambda$.

$$N_S = L(\lambda,\theta,\varphi) \cdot A_r \cdot \Omega_{FOV} \cdot \Delta\lambda \quad (2)$$

In terms of two-dimensional space (one-dimensional angle), eq. (2) can be translated to eq. (3), with $D_r$ being the aperture diameter of the telescope.

$$N_S = L(\lambda,\theta,\varphi) \cdot (\pi/4 \cdot D_r \cdot \theta_{FOV})^2 \cdot \Delta\lambda \quad (3)$$

The parameter that allows more control in this kind of link design (using an already-built telescope) is the detector area $d$, which has a strong relation with the optical resolution of the telescope. If $d$ is bigger than the PSF, then too much noise will reach the detector, decaying the SNR. If $d$ is smaller than the PSF, then many signal photons will be lost, which must be avoided in links as strongly limited by the received signal power as deep-space ones. Therefore, the detector size $d$ will determine the performance of the link, and ideally it should match the size of the PSF, which in turn should be as small as possible to minimize the FoV, thus the background noise, maximizing the SNR.

In IACTs, the PSF is mainly limited by telescope aberrations, which prevail over turbulence and diffraction effects. The PSF of MAGIC II was simulated by the authors to have a diameter of 31.7 mm [2], which agrees well with experimental measurements [9]. As this IACT has an aperture diameter of 17 m, its diffraction limit is in the order of tens of nm, five orders of magnitude below that value. The PSF affected by turbulence has been computed to be around 7.5 µrad for the worst case of 70º zenith angle at 1550 nm, over 3 orders of magnitude below the PSF. The MAGIC FoV for a single pixel is around 2 mrad, two orders of magnitude over usual values in FSOC receiver. CTA telescopes need to be characterized with their optical resolution to quantify this limitation when applied to a communication link, as well as to analyze possible PSF improvements to optimize their operation to FSOC. Next section will be devoted to this characterization.

*D. CTA telescopes simulations*

Telescopes with lower focal ratios (*f/D*) have stronger geometrical aberrations [10], which is the case for CTA telescopes. Furthermore, the use of spherical optics to reduce costs in the mirror manufacture, worsen the aberration behavior. For these reasons, ray-tracing simulations of CTA telescopes have been carried out to obtain their performance in terms of optical resolution. IACTs are designed to have a wide FoV, as they need to observe wide sections of the sky. Conversely, in FSOC, once the telescope is pointed towards the spacecraft, the radiation of interest will be received only in axis, hence FoV can be greatly reduced to maximize the SNR.

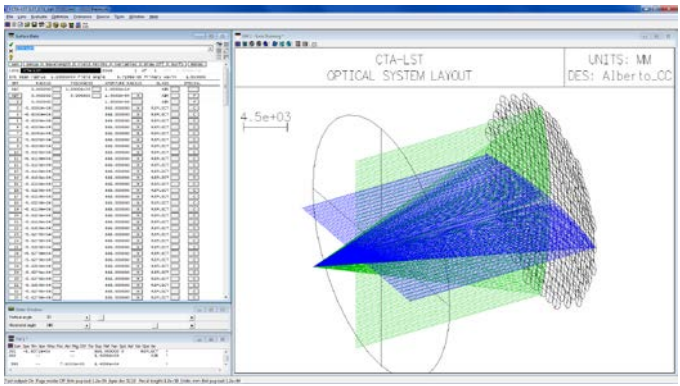

Fig. 4. Model in OSLO of the CTA-LST telescope.

CTA telescopes have been simulated in OSLO (see example of fig. 4) modelling their respective geometries and hexagonal spherical segments, each with its curvature radius according to its position in the primary mirror profile. In this study, aperture profiles are taken as they were designed for CTA: LST parabolic [11], MST [12] and SST [13] Davies-Cotton (MST-DC and SST-DC) [14], and MST and SST Schwarzschild-Couder (MST-SC and SST-SC) [15]. The latter one has two variations: the British-French GATE (GAmma-ray Telescope Elements): SST-SC GATE [16-17]; and the Italian ASTRI (Astrofisica con Specchi a Tecnologia Replicante Italiana): SST-SC ASTRI [18-19].

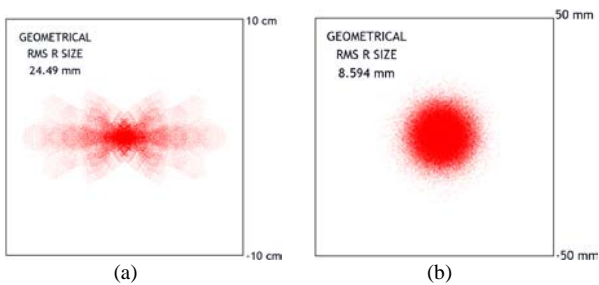

Fig. 5. (a) Spot diagram for non-optimized LST using ideal mirrors. (b) Spot diagram of the optimized telescope model with realistic mirrors.

As a first step, after computing the OSLO simulation assuming ideal mirrors, the following spot diagram radii were obtained for the segments of each telescope: 57.3 µm for SST-DC, 25.7 µm for MST-DC and 16.4 µm for LST (close to their diffraction limits, being 11.9 µm, 22 µm and 30.9 µm, respectively). Then, the final profiles were created by assembling all the segments according to the shape of each telescope. For each mirror (e.g. LST is made up by 198 different segments), it is necessary to compute its 3D position, its curvature radii and its tip/tilt angles: the 3D position of each mirror can be calculated according to each telescope profile, indicated in the previous paragraph; the curvature radii are calculated as an average value of the maximum and minimum values of the telescope profile in each segment; and as for the tip/tilt angles, initial values were calculated as a first approach using simple trigonometry.

The obtained spot diagrams were asymmetrical and needed to be optimized (fig. 5(a)). An iterative method for all the tip/tilt angles was applied to get an optimized and symmetrical spot diagram (fig. 5(b)). A damped least-squares method applied to an error function defined by the diameter of the PSF was applied in OSLO using its optimization capabilities. This is as realistic as using the actual tip/tilt actuators currently used in each IACTs to locally correct the alignment of each segment. The movements needed to perform this optimization were all below the actual capabilities of the Active Mirror Control (AMC) systems projected for CTA telescopes.

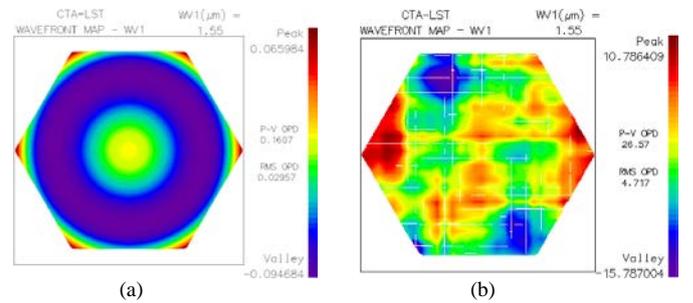

Fig. 6. (a) Wavefront of a LST mirror computed with an ideal model.
(b) Wavefront computed with a realistic distorted model.

When using real mirrors, the wavefront will be distorted beyond the ideal behavior. The performed OSLO models also took this effect into account simulating realistic prototypes of CTA mirrors from Japanese Sanko and Italian ASTRI [11, 21]. A series of random surface profiles were generated and superimposed over each ideal mirror to simulate this behavior (see the realistic model in fig. 6(b), compared to the ideal model in fig. 6(a)). Following similar statistical parameters as the experimental measurements from real samples, a different random surface was generated for each mirror. After this, more realistic spot diagrams were obtained.

The PSF size obtained with these OSLO simulations were as follows: 6.20 cm for LST, 3.42 cm for MST-DC, and 3.43 cm for SST-DC. Using their respective focal lengths, the FoV can be calculated as 2.24, 2.19 and 6.13 mrad, respectively for LST, MST-DC and SST-DC. For the SC configurations of MST and SST telescopes, not enough information was available to perform the OSLO simulations, therefore nominal design data were used instead, with FoV under 1 mrad [16-19]. These profiles allow achieving a better PSF due to the use of aspheric optics in order to reduce the pixel size, and hence the cost and size of the cameras, which can be built using smaller photodetectors (APDs instead of PMTs). This design constraint will be also an advantage for adapting the telescope to FSOC.

*E. Background noise*

To characterize the link in terms of SNR, the background noise power needs to be calculated. Eq. (2) must be solved for the worst scenario. The sky radiance $L(\lambda,\theta,\varphi)$ has been computed using MODTRAN, the usual reference for background noise computation in FSOC link budgets [21]. Following the suggestions from OLSG, the simulations used a minimum Sun-Earth-probe angle of 5° and a maximum observation zenith angle of 70°. The aerosol model applied was a maritime one, with 23-km visibility, taking into account a site like La Palma (Spain) for CTA North. Desert aerosol from Sahara also reaches Canary Islands, but the altitude of the observatory (2390 m) is usually beyond the influence of such an air mass.

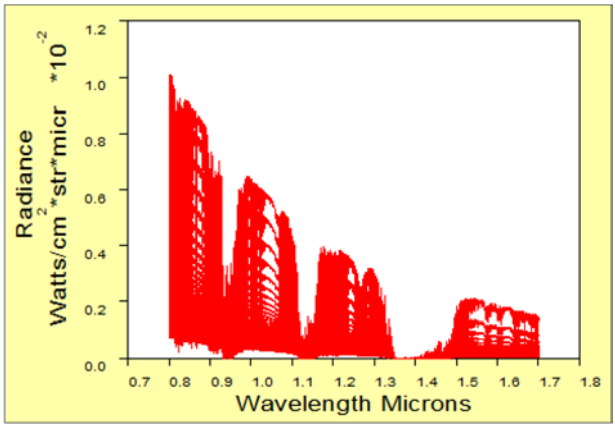

Fig. 7. MODTRAN Sky radiance for a maritime aerosol at La Palma (Spain) observatory. SZA 45°, azimuth 0°, observation zenith angles from 0° to 90°.

As an example, fig. 7 shows the MODTRAN simulated sky radiance for a solar zenith angle (SZA) of 45°, azimuthal angle of 0° and observation zenith angles from 0° to 90°. The highest curve is the one with equal observation and solar zenith angles. Sweeping the SZA for a wavelength of 1550 nm, the results in fig. 8 are obtained, after limiting the observation zenith angle to 70° and SEP angle to 5°. The average of all the maximum sky radiance values has been considered in this work as a worst case for daylight operation, giving a result of 430 µW/(cm²·srad·µm).

Next step is computing eq. (2) using the previous result. For noise power calculations, a spectral band $\Delta\lambda$ of 0.01 nm has been considered, although new filtering techniques can improve this figure in one order of magnitude [22-24]. For receivers, quantum efficiency has been taken as 75%, although 90% values have been achieved recently [25]. Background noise $N_S$ is therefore estimated as 70.69 nW for the LST as the worst case, 18.35 nW for MST-DC and 15.95 nW for SST-DC. SC configurations improve these values several orders of magnitude, thanks to their better optical resolution, which makes it possible to reduce the FoV: the $N_S$ for SST-SC ASTRI is 0.183 nW, being 0.052 nW for SST-SC GATE, and 0.041 nW for MST-SC.

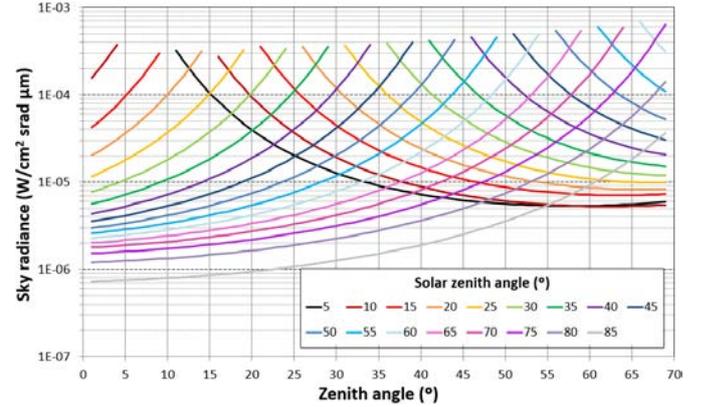

Fig. 8. Sky radiance as a function of zenith angle for different SZA angles at λ=1550 nm limited to observation zenith angle of 70º and SEP of 5º.

Night operation is also affected by background noise. In a link with another planet, its albedo is the highest contribution. For example, background noise power from Mars is computed with eq. (2) but using Mars irradiance $I_M$ (in W/cm²·µm) instead of the radiance $L$ due to the fact that Mars will be always included in the receiver's FoV. Eq. (4) was used to compute $I_M$ [26].

$$I_M = A_M \cdot (I_S/d_{M-S}^2) \cdot (R_M/d_{M-T})^2 \quad (4)$$

where $A_M$ is the 25% Mars albedo, $I_S$ = 28.7 mW/(cm²·µm) is the solar irradiance at 1 astronomical unit for 1500 nm, $d_{M-S}$ is the 1.52366 astronomical units Mars-Sun distance, $R_M$ is the 3390 km Mars radius, and $d_{M-T}$ is 68.6·10⁶ km, Mars-Earth distance in Mars opposition scenario, which is the case during nighttime. The background power $N_M$ by Mars albedo is then calculated with the eq. (5), resulting in 445 pW for LST, 120 pW for MST-DC and 7 pW for SST-DC.

$$N_M = I_M \cdot \pi (D/2)^2 \cdot \Delta\lambda \quad (5)$$

*F. Link budgets*

The link budget equation [21] applied to a downlink as the one simulated here is given by eq. (6), being $P_r$ (dBm) the power at the receiver, $P_t$ (dBm) the average transmitted power, $G_t$ (dB) the transmitter gain, $L_t$ (dB) the internal transmitter losses, $L_{tp}$ (dB) the transmitter pointing losses, $L_{fs}$ (dB) the free-space propagation losses, $L_{atm}$ (dB) the losses by atmospheric propagation, $G_r$ (dB) the receiver gain, $L_r$ (dB) the internal losses of the receiver and $L_{rp}$ (dB) the receiver pointing losses.

$$P_r = P_t + G_t - L_t - L_{tp} - L_{fs} - L_{atm} + G_r - L_r - L_{rp} \quad (6)$$

TABLE II. PARAMETERS USED IN THE LINK BUDGET CALCULATION.

|  | LEO | Moon | Lagrange L1 | Lagrange L2 | Mars opposit. | Mars conjunct. |
|---|---|---|---|---|---|---|
| Propagation distance (km) | $1.3·10^3$ | $384·10^3$ | $2·10^6$ | $2·10^6$ | $68.82·10^6$ | $400·10^6$ |
| Mean tx power (W) | 0.5 | 0.5 | 1 | 1 | 4 | 4 |
| Tx aperture diameter (cm) | 8 | 10.76 | 13.5 | 13.5 | 22 | 22 |
| Tx transmission (%) | 50 | 33 | 35 | 35 | 30.3 | 30.3 |
| Tx pointing losses (dB) | 0.11 | 0.31 | 0.08 | 0.08 | 0.05 | 0.05 |
| Bit rate (bit/s) | $10·10^9$ | $622·10^6$ | $120·10^6$ | $700·10^6$ | $260·10^6$ | $764·10^3$ |
| Scintillation losses (dB) | 2 | 1 | 2 | 2 | 0.2 | 0.2 |
| Atmospheric losses (%) | 86.2 | 90.3 | 86.2 | 86.2 | 95 | 95 |
| PPM symbols | (OOK) | 16 | 64 | 16 | 16 | 128 |
| Rx transmission (%) | 50 | 46.3 | 35 | 35 | 32.4 | 32.4 |
| Operation | Night and day | Night and day | Day | Night | Night | Day |

Eq. (6) has been used for each telescope and for several realistic scenarios: LEO (Low Earth Orbit), Lagrange points 1 and 2, Moon and Mars in conjunction and opposition (fig. 9). For the transmitter and other general parameters, the data from the scenario study suggested by the OLSG [20] was used (see table II). The rest of the link budget parameters were calculated from the simulations presented in the previous sections.

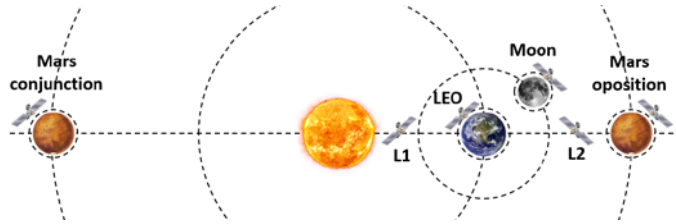

Fig. 9. Downlink scenarios simulated in this work.

As an example of operation, L1 scenario is shown in fig. 10. L1 is as a worst-case scenario regarding background noise. FoV and detection areas have been superimposed to the SNR in the plot for each CTA telescope. As can be seen, SNR is not dependent on the collection area, but on the FoV (the higher the FoV, the lower the SNR). It can be concluded that under high background, as it is the case of L1, LEO, Moon and Mars conjunction, it is the FoV rather than the aperture what determines the SNR. This can be used to compare the performance of CTA telescopes when applied to FSOC. According to this, in the fig. 10 CTA telescopes are shown in order of preference from right to left, being MST-SC the most advantageous CTA telescope for deep-space communications.

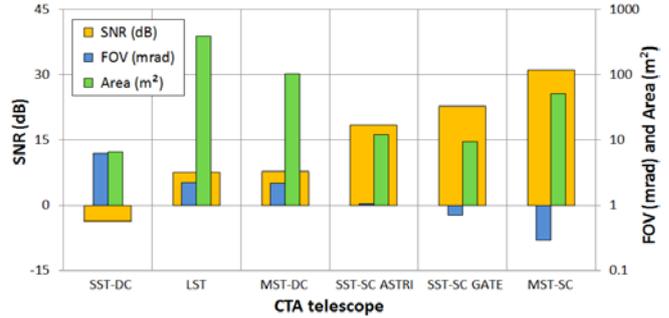

Fig. 10. Field of view (FoV) and aperture area compared with signal-to-noise ratio (SNR) for each CTA telescope in L1 scenario.

To check the ability of CTA telescopes to sustain FSOC links, in Table III, an overview of link budgets and worst-case SNR for each scenario is shown. 3 dB was selected as the link margin to overcome. Additionally, the condition of at least one photon per pulse per telescope was imposed, assuming the use of single-photon detectors, commonly used in quantum communications [27] and also recently demonstrated for deep-space lasercom links [28]. For each scenario, smaller or DC telescopes were first selected over SC to obtain the minimum SNR for being the simplest solutions. In nighttime and daytime scenarios, the worst-case (daytime) sky radiance values were used, and in night-only scenarios the Mars albedo background noise was used. Very occasional moon crossings in FoV have not been considered. Moon scenario is based on NASA's LLCD, L1 on ESA-NASA's SOHO mission and L2 on ESA's Euclid mission.

Receiver's gain is very important and it is the reason of higher SNR even for the smallest CTA telescopes in favorable scenarios (especially LEO and L2) where smaller telescopes would be enough. For example, $G_r$ difference in LEO between OLSG projected telescope and SST-DC is 17 dB.

As the Moon scenario is based on NASA's LLCD, the bitrate is variable between 40-622 Mbit/s (depending on the background and atmospheric conditions). Here, both the maximum bitrate and the worst-case background were used simultaneously, as a SST-DC (the simplest CTA solution)

TABLE III. LINK BUDGET AND SNR FOR EACH SCENARIO ASSUMING WORST-CASE BACKGROUND-NOISE LEVELS. IN EACH CASE, THE SIMPLEST TELESCOPE ALTERNATIVES WERE PREFERRED (INDICATED IN $G_r$ BETWEEN BRACKETS, AS WELL AS THE NUMBER OF ELEMENTS WHEN USING AN ARRAY OPERATION).

|  | LEO | Moon | Lagrange L1 | Lagrange L2 | Mars opposition | Mars conjunction |
|---|---|---|---|---|---|---|
| $P_t$ (dBm) | 26.99 | 26.99 | 30 | 30 | 36.02 | 36.02 |
| $L_t$ (dB) | -3.01 | -4.81 | -4.56 | -4.56 | -5.19 | -5.19 |
| $G_t$ (dB) | 104.20 | 106.77 | 108.74 | 108.74 | 112.98 | 112.98 |
| $L_{el}$ (dB) | -260.46 | -309.86 | -324.2 | -324.2 | -354.91 | -370.22 |
| $L_{atm}$ (dB) | -2.64 | -1.44 | -2.64 | -2.64 | -0.42 | -0.42 |
| $G_r$ (dB) | 135.29 (1×SST-DC) | 135.29 (1×SST-DC) | 147.36 (1×MST-DC) | 135.29 (1×SST-DC) | 147.36 (1×MST-DC) | 147.36 (3×MST-SC) |
| $L_r$ (dB) | -3.01 | -3.34 | -4.56 | -4.56 | -4.89 | -4.89 |
| $L_{rap}$ (dB) | -0.11 | -0.31 | -0.08 | -0.08 | -0.05 | -0.05 |
| $P_r$ (dBm) | -2.75 | -50.72 | -62 | -62 | -69.1 | -78.7 |
| SNR (dB) | 45.22 | 3.27 | 7.7 | 25.3 | 6.15 | 3.74 |

provides enough margin to close the link.

The L2 scenario is the most favorable one regarding background conditions (although the Mars opposition background were applied as a worst case), and using only one SST would allow improving the link. For example, instead of the 700 Mbit/s assumed by OLSG, a 3.7 Mbit/s link would still provide a SNR of 19.28 dB.

L1 is based on the same transmitter as L2. Hence, in order to adapt the link to the worse background conditions, the bitrate was reduced and the PPM order increased. This scenario suffers from a lot of background noise from the Sun, but one MST-DC could still have a link margin of 4.7 dB. If using SST-DC, an array of five telescopes would be necessary, as each one adds 3 dB, being mandatory at least one photon per pulse per telescope.

Mars opposition shows similar difficulties as L1 regarding background noise, adding 30 dB of losses because of the longer distance. Increasing the aperture diameter in this case is not useful, as Mars albedo is the only background source and it enters totally in the FoV, being in this case an irrelevant parameter. SST-DC cannot be used in spite of SNR>3 dB, because 0.2 photons per pulse would be received. MST-DC is the alternative, with a 3.15 dB link margin.

Mars in conjunction is the worst case scenario: the number of background photons per pulse is three orders of magnitude above signal photons. The only way to achieve a correct detection is using thousands of SST-DC, hundreds of MST-DC or tens of SST-SC. Only MST-SC in an array of 3 elements could fulfill the 3-dB link requirement.

TABLE IV. SNR IN dB FOR EACH TELESCOPE AND SCENARIO, INCLUDING BETWEEN BRACKETS THE No. OF TELESCOPES PER ARRAY TO OVERCOME 3 dB.

|  | LEO | Moon | L1 | L2 | Mars opp. | Mars conj. |
|---|---|---|---|---|---|---|
| SST-DC | 45.22 ✓ | 3.27 ✓ | -3.75 ✗ (5×) 3.2 | 25.30 ✓ | 6.15 ✓ | -35.88 ✗ (7750×) 3 |
| CTA-LST | 56.52 ✓ | 14.58 ✓ | 7.55 ✓ | 25.30 ✓ | 6.15 ✓ | -33.61 ✗ (4600×) 3 |
| MST-DC | 56.67 ✓ | 14.73 ✓ | 7.70 ✓ | 25.30 ✓ | 6.15 ✓ | -24.43 ✗ (560×) 3 |
| SST-SC ASTRI | 67.35 ✓ | 25.37 ✓ | 18.35 ✓ | 25.30 ✓ | 6.15 ✓ | -13.78 ✗ (48×) 3 |
| SST-SC GATE | 71.75 ✓ | 29.80 ✓ | 22.77 ✓ | 25.30 ✓ | 6.15 ✓ | -9.35 ✗ (18×) 3.2 |
| MST-SC | 80.07 ✓ | 50.03 ✓ | 31.09 ✓ | 25.30 ✓ | 6.15 ✓ | -1.04 ✗ (3×) 3.7 |

Table IV shows a summary of the worst-case SNR for each scenario and each CTA telescope when used as FSOC receiver. SNR is computed using the above described link budget and background noise simulations. Between brackets, the number of telescopes in an array configuration to close the link when a single telescope is not enough, is indicated. Only L1 with SST-DC and Mars in conjunction links are not fulfilled using single telescopes. The rest of the cases are feasible with any CTA telescope, proving that all the CTA telescopes could be used in FSOC links. In this study, the MST-SC was identified as the optimum solution and could serve as a ground station for all the scenarios except Mars in conjunction, where an array of 3 elements would be required.

*G. Cost*

Regarding costs, Table V shows a summary of the comparative costs between different ground stations. It can be seen that the cost of SST-SC is the lowest one, having an excellent performance compared with the DC configurations and making it possible to close the simulated links in every scenario except Mars conjunction. In general, CTA telescopes are cost-effective options when compared to FSOC telescopes and especially with astronomical telescopes. However, the FSOC telescopes include the adaptations for daylight operation and CTA telescopes should be also adapted. 1-m class telescope is estimated to need 1 M€ adaptation [20]. Scaling the costs for bigger apertures, CTA telescopes still would have lower cost than previously projected FSOC telescopes.

TABLE V. APPROXIMATED GROUND STATIONS COSTS ACCORDING TO THEIR TYPE (IACT, ASTRONOMICAL OR FSOC).

| Telescope | Type | Cost |
|---|---|---|
| CTA SST-SC | IACT | <0.5 M€[29] |
| CTA MST-DC | IACT | 1.6 M€[30] |
| CTA LST | IACT | 7.4 M€[30] |
| GTC / Keck | Astronomical | 100 M€[34] |
| Hobby-Eberly / SALT | Astronomical | 50 M€[31] |
| OLSG LEO | FSOC | 3.4 M€[20] |
| OLSG Moon | FSOC | 15.3 M€[20] |
| OLSG L1 | FSOC | 12.5 M€[20] |
| OLSG L2 | FSOC | 10.9 M€[20] |
| OLSG Mars | FSOC | 102.8 M€[20] |

*H. Additional proposed PSF improvements*

There is still an improvement margin for optical resolution in CTA telescopes: PSF size can be reduced by using certain techniques in order to employ just one single telescope dedicated to FSOC at each CTA site. As exposed previously, FoV in CTA telescopes ranges from 0.29 mrad (MST-SC) to 6.13 mrad (SST-DC), being 0.02 mrad the one assumed at OLSG proposals [20]. A brief list of possible adaptations is presented here in order to enhance optical performance in CTA primary-focus telescopes:

- New mirrors to prevent aberrations: all the CTA primary-focus telescopes use spherical optics to reduce costs, since a huge number of mirrors need to be built. Aspheric mirrors should be used instead to greatly improve the optical resolution, e.g. shaping a parabola in LST to approximate to the original structure profile. This would be very appropriate for FSOC, as parabolic mirrors ideally lack spherical aberration, and the coma would not be very harmful when using the narrow FoV needed in FSOC.

- Field corrector based on group of lenses or mirrors near the focal plane: they are also made by using aspheric optics, as in the case of COSTAR for Hubble telescope

[32], or the solution of Hobby-Eberly telescope, in which spherical aberration was corrected by using a group of 4 aspherical mirrors, reducing the PSF in three orders of magnitude [33]. FSOC case is also studied in [34] or [35], where a segmented spherical reflector is covering the 8.3-m central area of a 34-m NASA's Deep-Space-Network RF antenna, and a group of four 70-cm mirrors is used as a field corrector.

- Adaptive optics, as in the case of [36] or [37], where 8 additional dB could be gained when correcting atmospheric turbulence in a deep-space link. However, this solution should be studied to assess whether it is applicable or not for correcting the big aberrations of CTA telescopes.

## IV. CONCLUSIONS

Free-space lasercom holds the promise to alleviate the need of faster communications from deep-space. A key step for achieving this goal will be the development of a network of optical ground stations with very large apertures. However, such facilities will require big investments and new ideas are needed to minimize costs while maximizing the receiving apertures. The reutilization of astronomical facilities has been pointed out as a strategic action in this endeavor. A big number of Cherenkov telescopes will be built for CTA project in the next years. In this work, the feasibility of using CTA telescopes for deep-space FSOC has been explored. A deep analysis based on all the types of the projected CTA telescopes has been carried out. The reflectance of the CTA mirrors has been validated at 1550 nm by experimental spectral measurements, and a study of the limitations of FoV has been made, concluding that the geometrical aberrations are the limiting factor in CTA telescopes performance when applied to FSOC. A series of OSLO simulations have been carried out to retrieve the PSF of each telescope. Other simulations with MODTRAN and Matlab were made to obtain the optical link budget for realistic scenarios, considering worst cases.

With the only adaptation of replacing the Cherenkov camera by the lasercom equipment and its suitable refocusing by a few centimeters towards the focal plane, the possibility of using a MST-SC telescope for deep-space FSOC has been suggested. This telescope is able to reach over 6-dB SNR for every scenario except Mars in conjunction, where an array of 3 elements would be necessary to close the link. Three possible strategies have been suggested to enhance the optical performance of CTA primary-focus telescopes in order to improve their optical resolution to minimize their field-of-view and the received background noise: using aspherical mirrors, field correctors and adaptive optics. A brief discussion of the costs of Cherenkov telescopes compared with astronomical and communication telescopes was made. The conclusion is that the cost of optical ground stations based on Cherenkov telescopes would be lower than other dedicated FSOC telescope currently projected.

## ACKNOWLEDGMENT

Authors would like to thank: Juan Cabrero from INTA-ISDEFE (Spain) for performing some of the reflectance measurements; Purificación Munuera as the leader of the project at INSA (Spain) that was the basis for this study; MAGIC team at La Palma (Spain) for allowing access to the telescope facilities; CSIC Department of Metrology (Spain) for the reflectance measurements assistance; and Andreas Förster, responsible of mirrors testing at CTA, for the supplied samples.